\documentclass[12pt]{article}
\usepackage{graphicx}
\def \bea{\begin{eqnarray}}
\def \beq{\begin{equation}}

\def \eea{\end{eqnarray}}
\def \eeq{\end{equation}}

\def \ket#1{| #1 \rangle}

\def \od{\overline{D}^0}
\def \ok{\overline{K}^0}
\def \pp{\psi'}
\def \ppp{\psi''}
\def \s{\sqrt{2}}
\def \st{\sqrt{3}}
\def \sx{\sqrt{6}}

\begin{document}
\rightline{CLNS 04/1877}
\rightline{hep-ph/0405196}
\bigskip
\centerline {\bf STATUS OF $\ppp$ DECAYS TO CHARMLESS FINAL STATES
\footnote{Submitted to Physical Review D.}}
\bigskip
 
\centerline{Jonathan L. Rosner~\footnote{rosner@hep.uchicago.edu.  On leave
from Enrico Fermi Institute and Department of Physics,
University of Chicago, 5640 S. Ellis Avenue, Chicago, IL 60637}}
\centerline{\it Laboratory of Elementary Particle Physics}
\centerline{\it Cornell University, Ithaca, NY 14850}
\medskip
\centerline{(Received May 25, 2004)}
\medskip
 
\begin{quote}

The non-$D \bar D$ decays of the $\ppp = \psi(3770)$ resonance are discussed
and possibilities for further measurements are noted.  These decays can shed
light on S--D mixing, the ``missing'' $\psi' = \psi(3686)$ decays, a possible
discrepancy between the total and $D \bar D$ cross sections at the $\ppp$, and
rescattering effects contributing to enhanced $b \to s$ penguin amplitudes in
$B$ meson decays.  The importance is stressed of measurements (including the
$\psi''$ line shape) in states of definite G-parity and in inclusive charmless
final states such as $\eta' + X$ which are enhanced in charmless $B$ decays.

\end{quote}
\medskip

\centerline{PACS numbers:  13.25.Gv, 13.20.Gd, 14.40.Gx, 12.39.Jh}
\bigskip

\centerline{\bf I.  INTRODUCTION}
\bigskip

The $\psi'' \equiv \psi(3770)$ particle is the lowest-mass charmonium resonance
above $D \bar D$ threshold.\footnote{Numbers in parentheses will denote masses
of particles, in MeV$/c^2$.}  It is produced through virtual photons in $e^+
e^-$ collisions, and can serve as a well-defined source of charmed particle
pairs \cite{Eichten:1974af} in the same way that the $\Upsilon(4S)$ state,
lying just above $B \bar B$ threshold, is a good source of nonstrange $B$
mesons.  While discovered a number of years ago \cite{Rapidis:1977cv}, the
$\ppp$ is now being exploited through high-intensity studies under way at the
CLEO Detector at Cornell \cite{CLEOc} and at the BES Detector in China
\cite{Rong:2004ei}.  The couplings of the $\ppp$ to charmless states are of
interest for a number of reasons.

(1) $\ppp$ production and decays are sensitive to the mixing between S and D
waves in its wave function.  It is predominantly a $c \bar c(1^3D_1)$ state,%
\footnote{The spectroscopic shorthand is $n^{2S+1}L_J$, where $n=1,2,3,
\ldots$ is the radial quantum number; $S=0$ or 1 is the $c \bar c$ spin;
$L=S,P,D,\ldots~(l = 0, 1, 2, \ldots)$ is the orbital angular momentum, and
$J=0,1,2,\ldots$ is the total spin.} but contains important contributions
from mixing with the $2^3S_1$ level (of which $\psi' = \psi(3686)$ contains
the major share) as well as with possible other $^3S_1$ levels and $D \bar D$
continuum states \cite{Eichten:2002qv,Eichten:2004uh}.  This mixing can affect
not only $\ppp$ modes, but also $\psi'$ modes, suppressing some of them while
leading to contributions in $\ppp$ decays \cite{Rosner:2001nm} which may be
hard to see in those decays as a result of interference effects
\cite{Wang:2003hy,Yuan:2003hj,Wang:2003zx,Wang:2004qn,Wang:2004qg}.

(2) New measurements of the cross section $\sigma(e^+ e^- \to \ppp \to D \bar
D) \equiv \sigma(D \bar D)$ have recently been reported by the BES
\cite{Rong:2004ei}
and CLEO \cite{sigDD} Collaborations.  The CLEO measurement employs a
``double-tag'' method pioneered by the Mark III group in a previous study
\cite{Adler:1987as}.  Although the values appear higher than that of Mark III,
$\sigma(D \bar D)$ still seems less than the total cross section $\sigma(e^+
e^- \to \ppp \to \ldots) \equiv \sigma(\ppp)$, for which several groups have
reported measurements \cite{Partridge:1984kb,Peruzzi:1977ms,Schindler:1980ws,%
Bai:2001ct}.  If $\sigma(D \bar D)$ is really less than $\sigma(\ppp)$, this
would be a question of intrinsic interest and would provide an
estimate for rates for channels other than $D \bar D$ during the forthcoming
extensive accumulations of data by CLEO and BES-III at the $\ppp$ energy.

(3) The non-charm decays of $\ppp$, if appreciable, provide a possible
laboratory for the study of rescattering effects relevant to $B$ meson decays.
As one example, if the $\ppp$ decays to $D \bar D$ pairs which subsequently
re-annihilate into non-charmed final states, similar effects could be
responsible for enhanced penguin amplitudes (particularly in $b \to s$
transitions) in $B$ decays.  One particle whose enhanced production in both
exclusive and inclusive $B$ decays is not well understood is the $\eta'$.
It should be looked for in inclusive $\ppp$ decays.

The importance of a re-annihilation mechanism for possible decays of the $\ppp$
into non-charmed final states was stressed quite early \cite{Lipkin:1986av}.
Similar mechanisms are relevant not only to heavy quarkonium decays
\cite{Achasov:vh} but also to the decays of the $\phi$ meson into non-$K \bar
K$ states \cite{Achasov:1999tp}.  Non-charmed final states of the $\ppp$ were
discussed in two doctoral theses \cite{Zhu:1988,Majid:1993} based on Mark III
data.  However, no statistically significant signals were obtained.  Whereas
the total width of $\ppp$ is quoted \cite{PDG} as $\Gamma(\ppp) = 23.6 \pm 2.7$
MeV, partial widths to such known channels as $\gamma \chi_{cJ}~(J=1,2)$
and $J/\psi \pi \pi$ are expected not to exceed a few tens of keV, with a few
hundred keV expected for $\Gamma(\ppp \to \gamma \chi_{c0})$.  Thus any
significant non-$D \bar D$ branching ratio in excess of a percent or two must
come from as-yet-unseen hadronic channels or something more exotic.

I begin in Section II by reviewing the total and $D \bar D$ cross sections
at the $\ppp$.  Information from its leptonic and radiative decays (as well as
those of $\psi'$) is presented in Section III, while Section IV treats the
corresponding information available from $\ppp \to J/\psi \pi^+ \pi^-$.  An
important contribution to hadron production at the $\ppp$ energy comes from
the continuum process $e^+ e^- \to \gamma^* \to {\rm~light}~q \bar q
{\rm~pairs}$, treated in Section V.  The question of whether there is a
significant non-$D \bar D$ cross section is examined in Section VI.  If so,
charmless decays can illuminate certain classes of $B$ decays, as pointed out
in Section VII, where the particular utility of {\it inclusive} measurements
is stressed.  Section VIII concludes.  An Appendix contains details of a
model for re-annihilation of $D \bar D$ pairs into light quarks.
\bigskip

\centerline{\bf II.  CROSS SECTIONS AT THE $\ppp$ RESONANCE PEAK}
\bigskip

One can measure the cross section for $D \bar D$ production at the $\ppp$ by
comparing the rates for $e^+ e^- \to \ppp \to f_i + \ldots$ and $e^+ e^- \to
\ppp \to f_i \bar f_j$, where $f_i$ and $f_j$ are final states in $D$ decay.
Unknown branching ratios can be determined, but one must have good knowledge of
detector efficiency.  This method was first used by the Mark III Collaboration
\cite{Adler:1987as} to determine $\sigma(D \bar D) =(5.0 \pm 0.5)$ nb, based on
an integrated luminosity $\int {\cal L} dt = (9.56 \pm 0.48)$ pb$^{-1}$.

The CLEO Collaboration has recently measured $\sigma(D \bar D)$ using this
same double-tag method but with $\int {\cal L} dt \simeq 57$ pb$^{-1}$
\cite{sigDD}.  The values are compared with those from Mark III and from a
single-tag measurement by the BES Collaboration \cite{Rong:2004ei} (with $\int
{\cal L} dt = \\ 17.7$ pb$^{-1}$) in Table \ref{tab:sigDD}.

\begin{table}
\caption{Comparison of cross sections $\sigma(D \bar D) \equiv \sigma(e^+ e^-
\to \ppp \to D \bar D)$, in nb.
\label{tab:sigDD}}
\begin{center}
\begin{tabular}{l c c c} \hline \hline
Collaboration & $\sigma(D^+ D^-)$ & $\sigma(D^0 \bar D^0)$ & $\sigma(D \bar D)$
 \\ \hline
BES-II$^a$ \cite{Rong:2004ei} & $2.52 \pm 0.07 \pm 0.24$ & $3.26 \pm 0.09
 \pm 0.25$ & $5.78 \pm 0.11 \pm 0.45$ \\
CLEO$^a$ \cite{sigDD} & $2.59 \pm 0.11 \pm 0.11$ & $3.47 \pm 0.07 \pm 0.15$
 & $6.06 \pm 0.13 \pm 0.23$ \\
Mark III \cite{Adler:1987as} & $2.1 \pm 0.3$ & $2.9 \pm 0.4$ & $5.0 \pm 0.5$ \\
\hline \hline
$^a$ Preliminary. & & & \\
\end{tabular}
\end{center}
\end{table}

The ratios $\sigma(D^+ D^-)/\sigma(D^0 \bar D^0)$ are consistent at present
with the ratio of kinematic factors $(p^*_{+-}/p^*_{00})^3 = 0.685$ appropriate
for the P-wave decay $\ppp \to D \bar D$ (where $p^*$ denotes the magnitude of
the center-of-mass [c.m.] 3-momentum).  Coulomb and other
final-state-interaction effects
can alter this ratio and lead to its dependence on energy
\cite{Voloshin:2004nu}, but these phenomena remain to be studied.

The values in Table \ref{tab:sigDD} are to be compared with those for the
total cross section $\sigma(\ppp)$ in Table \ref{tab:sig}.  It is possible that
$\sigma(D \bar D)$ falls short by one or more nb from the total cross section
$\sigma(\ppp)$, but the difference is not statistically significant.  Improved
measurements of both quantities by the {\it same} experiment will be needed to
resolve the question.  In Section VII we will take an illustrative example in
which this difference, taken to be 18\% of $\sigma(\ppp)$, is ascribed to
re-annihilation of $D \bar D$ into light-quark states.

\begin{table}
\caption{Comparison of total cross sections $\sigma(\ppp) \equiv \sigma(e^+ e^-
\to \ppp \to \ldots)$, in nb.
\label{tab:sig}}
\begin{center}
\begin{tabular}{l c} \hline \hline
Collaboration & $\sigma(\ppp)$ \\ \hline
Crystal Ball \cite{Partridge:1984kb}  & $6.7 \pm 0.9$  \\
Lead-Glass Wall \cite{Peruzzi:1977ms} & $10.3 \pm 1.6$ \\
Mark II \cite{Schindler:1980ws}       & $9.3 \pm 1.4$  \\
BES$^a$ \cite{Bai:2001ct}            & $7.7 \pm 1.1$  \\ \hline
Average                               & $7.9 \pm 0.6$  \\ \hline \hline
$^a$ Estimate based on fit (see Sec.\ VI). & \\
\end{tabular}
\end{center}
\end{table}

\bigskip

\centerline{\bf III.  INFORMATION FROM LEPTONIC AND RADIATIVE DECAYS}
\bigskip

\begin{figure}
\begin{center}
\includegraphics[height=6.5in]{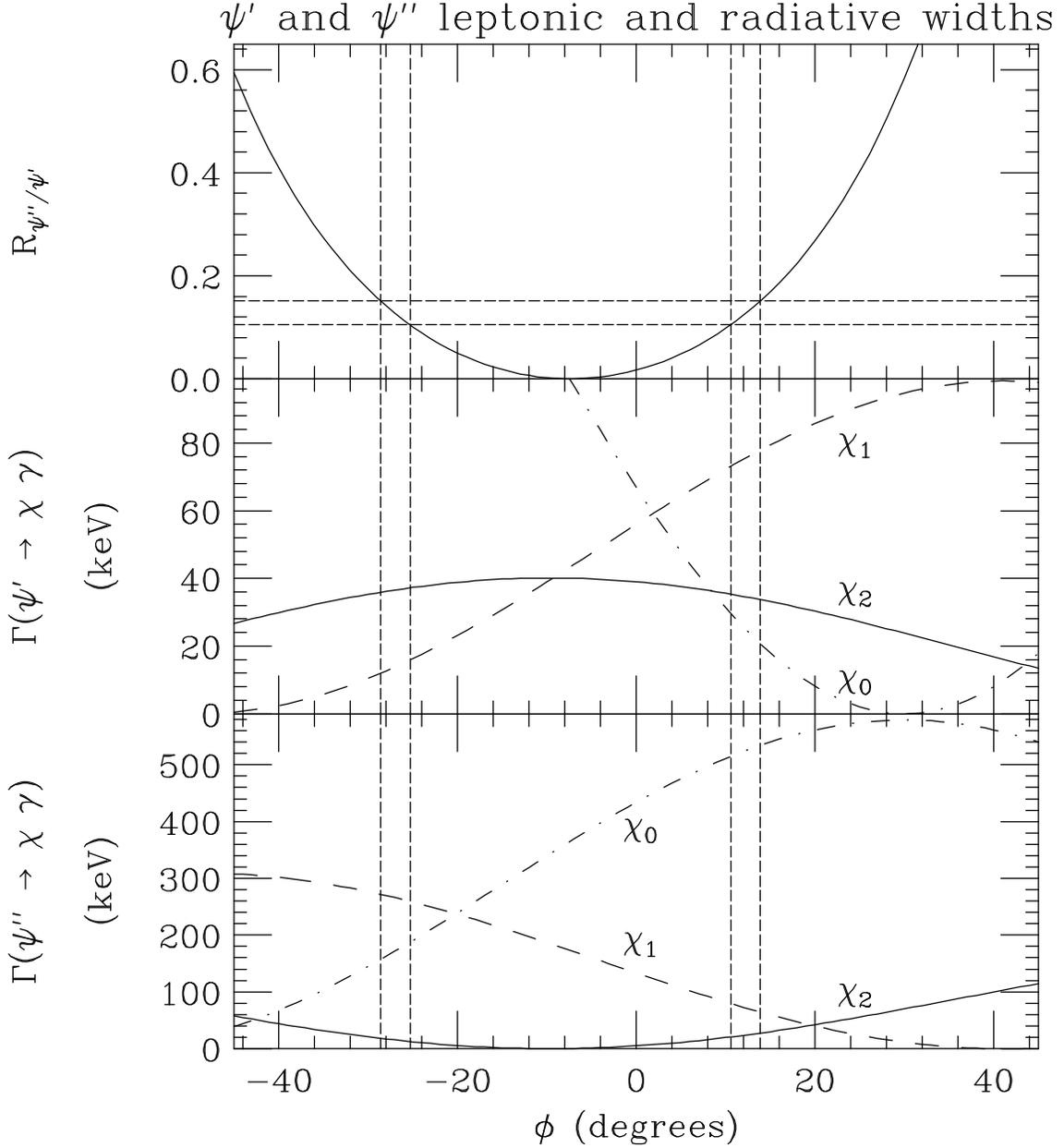}
\end{center}
\caption{Sensitivity of scaled leptonic width ratio $R_{\ppp/\psi'}$ and
partial widths $\Gamma(\psi',\ppp \to \chi \gamma)$ to mixing angle $\phi$.
Horizontal lines in top panel denote $\pm 1 \sigma$ limits on $R_{\ppp/\psi'}$,
and are projected onto the $\phi$ axis with vertical bands.  In middle and
bottom panels solid, dashed, and dash-dotted curves denote partial widths to
$\gamma \chi_{c2}$, $\gamma \chi_{c1}$, and $\gamma \chi_{c0}$, respectively.
\label{fig:psimix}}
\end{figure}

A simple model of S--D wave mixing for the $\psi'$ and $\ppp$ is to write
\begin{equation}\label{eqn:mix}
\ppp = \cos \phi \ket{1 ^3D_1} + \sin \phi \ket{2 ^3S_1}~~,~~~
\psi' = -\sin \phi \ket{1 ^3D_1} + \cos \phi \ket{2 ^3S_1}~~.
\end{equation}
The ratio $R_{\ppp/\psi'}$ of leptonic widths (scaled by factors of $M^2$) and
the partial widths $\Gamma(\psi' \to \chi \gamma)$ and $\Gamma(\ppp \to \chi
\gamma)$ may then be calculated as functions of $\phi$ \cite{Rosner:2001nm,%
Kuang:2002hz}.  Specifically, it was found in Ref.\ \cite{Rosner:2001nm} that
\begin{equation}
R_{\ppp/\psi'} \equiv \frac{M_{\ppp}^2 \Gamma(\ppp \to e^+ e^-)}
{M_{\pp}^2 \Gamma(\pp \to e^+ e^-)}
= \left| \frac{0.734 \sin \phi + 0.095 \cos \phi}
              {0.734 \cos \phi - 0.095 \sin \phi} \right|^2
= 0.128 \pm 0.023~~,
\end{equation}
while
\begin{equation}
\Gamma(\ppp \to \gamma \chi_{c0}) = 145~{\rm keV} \cos^2 \phi
 (1.73 + \tan \phi)^2~~,
\end{equation}
\begin{equation}
\Gamma(\ppp \to \gamma \chi_{c1}) = 176~{\rm keV} \cos^2 \phi
 (-0.87 + \tan \phi)^2~~,
\end{equation}
\begin{equation}
\Gamma(\ppp \to \gamma \chi_{c2}) = 167~{\rm keV} \cos^2 \phi
 (0.17 + \tan \phi)^2~~,
\end{equation}
and
\begin{equation}
\Gamma(\pp \to \gamma \chi_{c0}) = 67~{\rm keV} \cos^2 \phi
 (1 - 1.73 \tan \phi)^2~~,
\end{equation}
\begin{equation}
\Gamma(\pp \to \gamma \chi_{c1}) = 56~{\rm keV} \cos^2 \phi
 (1 + 0.87 \tan \phi)^2~~,
\end{equation}
\begin{equation}
\Gamma(\pp \to \gamma \chi_{c2}) = 39~{\rm keV} \cos^2 \phi
 (1 - 0.17 \tan \phi)^2~~.
\end{equation}
These quantities are plotted as functions of $\phi$ in Fig.\ \ref{fig:psimix}.

The observed ratio $R_{\ppp/\psi'}$ agrees with predictions only for
$\phi = (12 \pm 2)^\circ$ or $(-27 \pm 2)^\circ$, as shown by the vertical
bands in Fig.\ \ref{fig:psimix}.  Only the solution with $\phi = (12 \pm
2)^\circ$ is remotely consistent with the observed partial widths
\cite{PDG} $\Gamma(\psi' \to \gamma \chi_{cJ}) = 20$--30 keV.  This range of
$\phi$ favors the decay $\ppp \to \gamma \chi_{c0}$ over $\ppp \to \gamma
\chi_{c1,2}$ by a substantial amount.  The choice $\phi = (12 \pm 2)^\circ$
also is favored by the comparison of $\pp$ and $\ppp$ decays to $J/\psi
\pi^+ \pi^-$.  With the choice $\phi = (-27 \pm 2)^\circ$,
a larger rate would be predicted for $\ppp \to J/\psi \pi^+ \pi^-$ than for
$\pp \to J/\psi \pi^+ \pi^-$, in conflict with experiment \cite{Kuang:ub}.
It has recently been argued \cite{Liu:2004un} that the mixing could be
larger, $|\phi| \simeq 40^\circ$, but this conclusion depends on specific
production models for the $\ppp$ in inclusive $e^+ e^-$ annihilations and in
$B$ decays.

The prospects for observation of $\ppp \to \gamma \chi_{cJ}$ have been
greatly improved with the accumulation of the recent data sample of $\int
{\cal L} dt \simeq 57$ pb$^{-1}$ in the CLEO-c detector \cite{sigDD}.
With this sample and $\sigma(\ppp) \ge 6$ nb one should see
several events in the cascade $\ppp \to \gamma \chi_{c1} \to \gamma \gamma
J/\psi \to \gamma \gamma \ell^+ \ell^-$.  The inclusive signal in $\ppp \to
\gamma \chi_{c0}$ will not be statistics-limited.  All predictions of
branching ratios lie in the 1--2\% range.

It is important to consider coupling to open $D \bar D$ channels and mixing
schemes that are more general than Eq.\ (\ref{eqn:mix}) when
predicting radiative decay widths \cite{Eichten:2004uh}.  Table \ref{tab:pw}
compares partial widths predicted in one such scheme with those
depicted in Fig.\ \ref{fig:psimix}.  In Ref.\ \cite{Eichten:2004uh} the
$\ppp$ is composed of only 52\% $ c \bar c$; the remainder of its wave
function contains additional light quark-antiquark pairs, {\it e.g.}, in the
form of the open $D \bar D$ channel. Thus the results of Fig.\ \ref{fig:psimix}
do represent some oversimplification.

\begin{table}
\caption{Partial widths in keV predicted in Ref.\ \cite{Eichten:2004uh}
without (a) or with (b) couplings to open channels and in Ref.\
\cite{Rosner:2001nm}.  $M(\ppp) = 3772$ MeV/$c^2$ is taken in accord
with the fit of Sec.\ VI; the nominal mass quoted in Ref.\ \cite{PDG} is
$3769.9 \pm 2.5$ MeV/$c^2$. \label{tab:pw}}
\begin{center}
\begin{tabular}{c c c c c} \hline \hline
$\psi''$& $E_\gamma$ & \multicolumn{2}{c}{Ref.\ \cite{Eichten:2004uh}} & Ref.\
\cite{Rosner:2001nm} \\
decay   &  (MeV)  & (a) & (b) & ($\phi = 12 \pm 2^\circ$) \\ \hline
$\gamma \chi_{c2}$ & 210 & 3.2 & 3.9 & $24 \pm 4$ \\
$\gamma \chi_{c1}$ & 252 & 183 & 59  & $73 \pm 9$ \\
$\gamma \chi_{c0}$ & 340 & 254 & 225 & $523 \pm 12$ \\ \hline \hline
\end{tabular}
\end{center}
\end{table}

For an exclusive decay involving $\chi_{c1}$ suppose that $\Gamma(\psi'' \to
\gamma \chi_{c1}) = 59$ keV and use the tabulated branching ratios \cite{PDG}
${\cal B}(\chi_{c1} \to \gamma J/\psi) =(31.6 \pm 1.2)\%$, ${\cal B}(J/\psi \to
\ell^+ \ell^-) = (5.9 \pm 0.1)\%$ ($\ell = e$ or $\mu$).  With an efficiency of
1/2 for each shower or charged track\footnote{This is a conservative
estimate; the CLEO-c detector can probably do considerably better.}
one expects to see two events.

The number of $\ppp \to \gamma \chi_{c0}$ events expected in the current CLEO-c
sample of $\simeq 57$ pb$^{-1}$ may be estimated as follows.  Suppose the cross
section for $\ppp$ production is at least 6 nb.  Assume a branching ratio of
at least a percent and a photon detection efficiency of at least 50\%.  Then
one expects at least $56000 \times 6 \times 0.01 \times 0.5 = 1680$ events
containing a monochromatic photon with energy 340 MeV.

The Mark III collaboration \cite{Zhu:1988} reported some marginal signals
for $\ppp$ radiative decays (quoted in Ref.\ \cite{Rosner:2001nm}), whose
partial widths we now adjust for the ratio of the Mark III total cross section
\cite{Adler:1987as} $\sigma(\ppp) = 5.0 \pm 0.5$ nb and our average of
$\sigma(\ppp) = 7.9 \pm 0.6$ nb:
\begin{equation}
\Gamma(\ppp \to \gamma \chi_{c0}) = (320 \pm 120)~{\rm keV}~,
\end{equation}
\begin{equation}
\Gamma(\ppp \to \gamma \chi_{c1}) = (280 \pm 100)~{\rm keV}~,
\end{equation}
with an upper limit
\begin{equation}
\Gamma(\ppp \to \gamma \chi_{c2}) \le 330~{\rm keV}~(90\%~{\rm c.l.})~.
\end{equation}
The partial widths predicted in Table \ref{tab:pw} imply that the signal
for $\ppp \to \gamma \chi_{c0}$ could be genuine, but that for $\ppp \to \gamma
\chi_{c1}$ is less likely to be so.
\bigskip

\centerline{\bf IV.  INFORMATION FROM $\ppp \to J/\psi \pi^+ \pi^-$}
\bigskip

The rate for the decay $\ppp \to J/\psi \pi^+ \pi^-$ was originally estimated
by Kuang and Yan \cite{Kuang:ub} using a QCD multipole expansion and assuming
the $\ppp$ to be a pure $^3D_1$ state.  The inclusion of mixing and comparison
with experimental results imply that the intrinsic $^3D_1 \to J/\psi \pi^+
\pi^-$ amplitude cannot be neglected but is not as large as a free-gluon
approximation would predict.

An early Mark III result reported in Ref.\ \cite{Zhu:1988} found $\sigma(\ppp)
{\cal B}(\ppp \to J/\psi \pi^+ \pi^-) = (1.2 \pm 0.5 \pm 0.2) \times 10^{-2}$
nb, implying ${\cal B}(\ppp \to J/\psi \pi^+ \pi^-) = (0.15 \pm 0.07)\%$.
This result is compared with others in Table \ref{tab:psipipi}.  The average
(not including information from the CLEO upper limit)
is ${\cal B}(\ppp \to J/\psi \pi^+ \pi^-) = (0.18 \pm 0.06)\%$, corresponding
to a partial width of $43 \pm 14$ keV.  Adding another 50\% for the
$\ppp \to J/\psi \pi^0 \pi^0$ mode, one finds $\Gamma(\ppp \to J/\psi \pi \pi)
= (64 \pm 21)$ keV, or at most about 100 keV.

\begin{table}
\caption{Comparison of experimental branching ratios ${\cal B}(\ppp \to J/\psi
\pi^+ \pi^-)$, in percent.
\label{tab:psipipi}}
\begin{center}
\begin{tabular}{l c} \hline \hline
Collaboration & ${\cal B}(\ppp \to J/\psi \pi^+ \pi^-)$ \\ \hline
Mark II \cite{Zhu:1988} & $0.15 \pm 0.07$  \\
BES \cite{Bai:2003hv}   & $0.34 \pm 0.14 \pm 0.08$ \\ 
Mark II -- BES average & $0.18 \pm 0.06$  \\
CLEO \cite{Skwarnicki:2003wn}& $< 0.26~(90\%~{\rm c.l.})$ \\ \hline \hline
\end{tabular}
\end{center}
\end{table}

Kuang and Yan \cite{Kuang:ub} predicted $\Gamma(\ppp \to J/\psi \pi^+ \pi^-) =
107$ keV for a free-gluon estimate of $^3D_1 \to J/\psi \pi^+ \pi^-$ (based on
the observed $\psi' \to J/\psi \pi^+ \pi^-$ rate) and $\Gamma(\ppp \to J/\psi
\pi^+ \pi^-) = 20$ keV if $\Gamma(^3D_1 \to J/\psi \pi^+ \pi^-)$ were reduced
by a factor of 3 from a free-gluon estimate.  (This estimate is lower than
107/3 because of the interplay of S-wave and D-wave contributions to the $\ppp$
decay.)  This may have implications for the search for $\Upsilon(1D) \to
\Upsilon(1S) \pi^+ \pi^-$.  A recent CLEO upper limit \cite{Bonvicini:2004}
for $\Upsilon(1D) \to \Upsilon(1S) \pi^+ \pi^-$ lies about a factor of 7 below
the Kuang-Yan \cite{Kuang:ub} free-gluon prediction.  Mixing in the
$\Upsilon(1D)$ may be different from that in $\ppp$, however.

A further complication in analysis of the $\ppp \to J/\psi \pi^+ \pi^-$
partial width arises from the tail of the $\pp$, whose contribution is
non-negligible at the $\ppp$ mass as a result of the large branching ratio for
$\pp \to J/\psi \pi^+ \pi^-$.  A thorough analysis of this effect probably
requires measurement of the energy dependence of the apparent $\ppp \to J/\psi
\pi^+ \pi^-$ signal.
\bigskip

\centerline{\bf V.  CONTINUUM EXPECTATIONS}
\bigskip

The total cross section for $e^+ e^- \to \ppp$, whether it is 6, 8, or 10 nb,
is by no means the only contribution to hadron production at a c.m. energy
of 3770 MeV.  One expects hadron production from $e^+ e^- \to q \bar q$
$(q=u,d,s)$ to account for
\begin{eqnarray}
\sigma(e^+ e^- \to q \bar q) & = & N_c \left[ \left(\frac{2}{3}
\right)^2 + \left(-\frac{1}{3} \right)^2 + \left(-\frac{1}{3} \right)^2 \right]
\sigma(e^+ e^- \to \mu^+ \mu^-) \left[ 1 + \frac{\alpha_s}{\pi}
+ \ldots \right]~~, \nonumber \\
& = & 2 ~\sigma(e^+ e^- \to \mu^+ \mu^-) \left[ 1 + {\rm QCD~ correction}
\right]~~, \label{eqn:R}
\end{eqnarray}
where $N_c = 3$ is the number of quark colors and (neglecting the muon mass)
\begin{equation}
\sigma(e^+ e^- \to \mu^+ \mu^-) = \frac{4 \pi \alpha^2}{3s}
= 6.1~{\rm nb}
\end{equation}
for $s \equiv E_{\rm c.m.}^2 = (3770~{\rm MeV})^2$.  This contribution will be
referred to as {\it continuum}.  In addition $\tau^+ \tau^-$ pair production
would account for
\begin{equation}
\sigma(e^+ e^- \to \tau^+ \tau^-) = \left( 1 - \frac{4 m_\tau^2}
{s} \right)^{1/2} \left( 1 + \frac{2 m_\tau^2}{s} \right)
\sigma(e^+ e^- \to \mu^+ \mu^-)~~~
\end{equation}
or about 2.9 nb if initial-state-radiation effects were neglected.  Such
effects will change the observed cross section.  The separation of
$\tau^+ \tau^-$ from $q \bar q$ final states requires good understanding of
detector sensitivities and $q \bar q$ fragmentation.

The couplings of virtual photons to two pseudoscalar mesons $P$ or one
pseudoscalar and one vector $V$ can be evaluated straightforwardly
\cite{Haber:1985cv,Kopke:1988cs}.  They are proportional to Tr($Q[P_1,P_2]$)
or Tr($Q\{P,V\}$), where $Q = {\rm Diag} (2/3,-1/3,-1/3)$ and [for one
$(\eta,\eta')$ mixing scheme which fits other data well
\cite{Kawarabayashi:1980dp}]

\begin{equation}
P \equiv \left[ \begin{array}{c c c} \frac{\pi^0}{\s} + \frac{\eta}{\st}
+ \frac{\eta'}{\sx} & \pi^+ & K^+ \cr \pi^- & - \frac{\pi^0}{\s}
+ \frac{\eta}{\st} + \frac{\eta'}{\sx} & K^0 \cr
K^- & \ok & - \frac{\eta}{\st} + 2 \frac{\eta'}{\sx}\end{array} \right]~~,~~~
\end{equation}

\begin{equation}
V \equiv \left[ \begin{array}{c c c} \frac{\rho^0}{\s} + \frac{\omega}{\s} &
\rho^+ & K^{*+} \cr \rho^- & - \frac{\rho^0}{\s} + \frac{\omega}{\s} &
K^{*0} \cr K^{*-} & \overline{K}^{*0} & \phi \end{array} \right]~~~.
\end{equation}

These couplings lead to the characteristic continuum ($\gamma^*$) production
ratios:
$$
\pi^+ \pi^-:~K^+ K^-:~ K^0 \ok = 1:~1:~0~~;
$$
\begin{equation}
\omega \pi^0:\rho \eta:K^{*0} \ok:\phi \eta:\rho \pi:K^{*+} K^-:
\omega \eta:\phi \pi^0 = 1:\frac23:\frac49:\frac{4}{27}:\frac13:\frac19:
\frac{2}{27}:0~.
\end{equation}
Here I have neglected a small admixture of nonstrange quarks in the $\phi$
responsible for its $\rho \pi$ decay.
The contribution of the isovector photon $(G=+)$ dominates: $\sigma(2 \pi+ 4
\pi+ \ldots) = 9 \sigma (3 \pi+ 5 \pi+ \ldots)$.  Thus one has several
signatures of continuum production which can be examined at a single
energy, e.g., at the $\ppp$ peak.  Of course, a better way to study continuum
contributions is to change the c.m.\ energy to one where resonance production
cannot contribute.  The CLEO Collaboration has done this, studying hadron
production at a c.m.\ energy of 3670 MeV with a sample of 21 pb$^{-1}$
\cite{Asner:2004yu}, and results are currently being analyzed.
\bigskip

\centerline{\bf VI.  IS THERE A SIGNIFICANT NON-$D \bar D$ CROSS SECTION?}
\bigskip

At most 600 keV of the $\ppp$ total width of $23.6 \pm 2.7$ MeV is due to
radiative decays, and perhaps as much as another 100 keV is due to $J/\psi \pi
\pi$ decays.  Along with the predominant $D \bar D$ decays, are these
contributions enough to account for the total $\ppp$ width?

\begin{figure}
\includegraphics[height=6.15in]{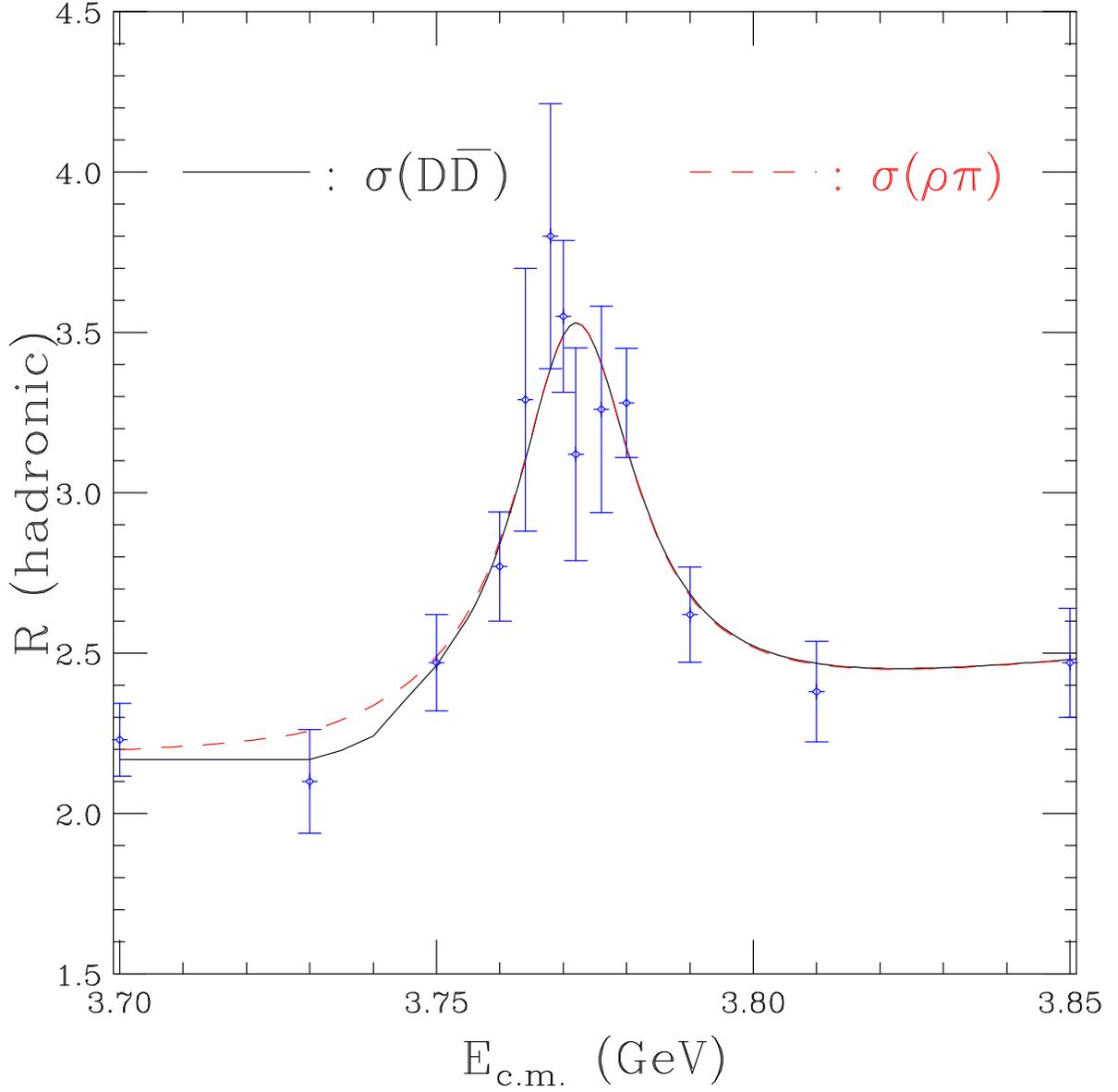}
\caption{Fit to the $\ppp$ peak in BES data \cite{Bai:2001ct}.  Solid line
denotes expected line shape for a $D \bar D$ final state, incorporating
appropriate centrifugal barrier terms, while dashed line denotes expected
line shape for $\rho \pi$ final state.
\label{fig:sigppp}}
\end{figure}

In Fig.\ \ref{fig:sigppp} the BES data \cite{Bai:2001ct} on $R = \sigma(e^+ e^-
\to {\rm hadrons})/\sigma(e^+ e^- \to \mu^+ \mu^-)$ are displayed, along with
the results of a fit to the resonance shape using conventional Blatt-Weisskopf
angular momentum barrier factors \cite{BW,VonHippel:fg}.  The fit obtains
$\sigma_{\rm pk} = 7.7 \pm 1.1$ nb, with other central values $M = 3772$ MeV,
$\Gamma = 23.2$ MeV, and $R_{\rm bg} = 2.17 + 2.36(E_{\rm c.m.} - 3.73~{\rm
GeV})\theta(E_{\rm c.m.} - 3.73~{\rm GeV})$, where the threshold energy of
3.73 GeV is held fixed in the fit \cite{TSPC}.

BES measurements in the energy range near the $\ppp$ resonance are consistent
with the expectation (\ref{eqn:R}).  In the c.m.\ energy range 2--3 GeV the
value $R = 2.26 \pm 0.14$ is obtained \cite{Rong:2004ei}.  This would imply
$\sigma(e^+ e^- \to {\rm hadrons}) \simeq 13.8 \pm 0.9$ nb at 3770 MeV,
quite a bit more than $\sigma(\ppp)$.  Thus inferring $\sigma(\ppp)$ by
subtracting known processes such as continuum from a cross section measured at
a single energy may carry large systematic errors.

Taking as an illustrative value $\sigma(D \bar D) \le 6.5$ nb and comparing it
with the overall average of $\sigma(\ppp) = 7.9$ nb in Table \ref{tab:sig}, one
is invited to consider how to account for a deficit of 1.4 nb, or 18\% of the
total.  While this quantity is not statistically significant, it is interesting
to speculate on possible sources pending (1) a scan of the $\ppp$ peak to
measure $\sigma(\ppp)$ more accurately and (2) reduction of the error on
$\sigma(D \bar D)$.
\bigskip

\centerline{\bf VII.  POTENTIAL INFORMATION FROM CHARMLESS MODES}
\bigskip

The possibilities for detecting {\it individual} charmless decay modes of the
$\ppp$ were raised, for example, in Refs.\ \cite{Rosner:2001nm} and
\cite{Achasov:vh}.  Here I stress that more inclusive measurements at
the $\ppp$ also may be of use.

\begin{figure}
\includegraphics[height=6.2in]{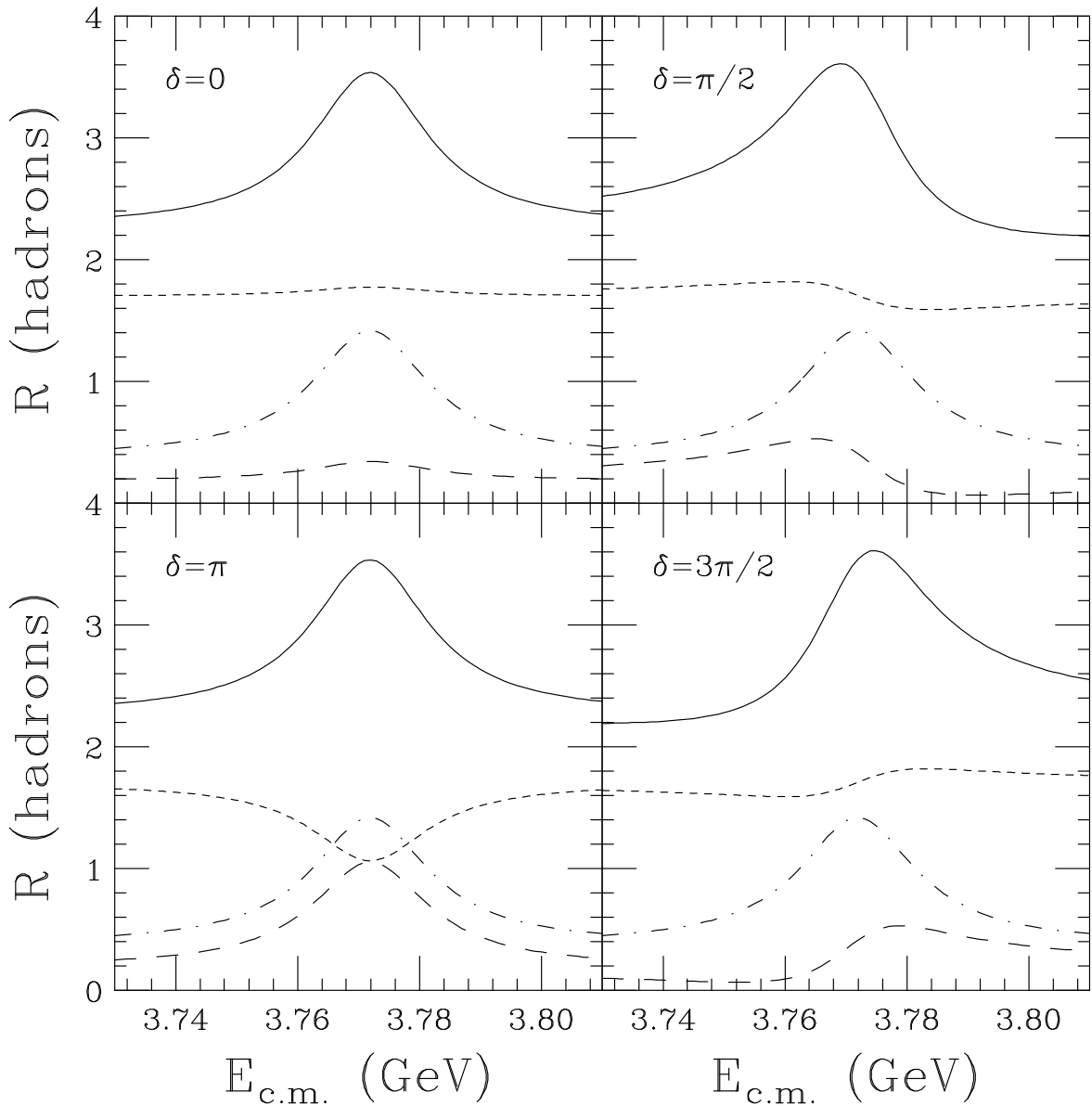}
\caption{Contributions to $R$ in the vicinity of the $\ppp$ resonance energy.
Solid curves: total, constrained to have a value of 3.53 at $M(\ppp) = 3.772$
GeV/$c^2$.
Short-dashed curves: $I=1$ continuum interfering with $I=1$ contribution from
$D \bar D$ reannihilation.  Long-dashed curves: $I=0$ non-strange continuum
interfering with $I=0$ nonstrange contribution from $D \bar D$ reannihilation.
Dot-dashed curves:  $D \bar D$ resonance contribution, taken to contribute
82\% of resonance peak cross section, plus $s \bar s$ continuum.
\label{fig:rall}}
\end{figure}

Consider a model in which the re-annihilation of charmed quarks in $D^0 \bar
D^0$ and $D^+ D^-$ into states containing $u,~d,~s$ accounts
for the difference between $\sigma(D \bar D)$ and $\sigma(\ppp)$.  The
possibility of such re-annihilation was considered some time ago
\cite{Lipkin:1986av} both as a source of non-$D \bar D$ decays of the $\ppp$
and as a possible source of non-$B \bar B$ decays of the $\Upsilon(4S)$.
The latter do not appear to occur at any level above a few percent
\cite{nonBB}.  As an illustration, we present in Fig.\ \ref{fig:rall}
the case in which such re-annihilation accounts for 18\% of the peak $R$
value at $M(\ppp) = 3772$ MeV/$c^2$.  A relative phase $\delta$ between
the reannihilation amplitude and the continuum was defined in such a way
that $\delta = 0$ corresponds to constructive interference at the resonance
peak.  Details of this model are given in the Appendix.

Several features of this model are worth noting.

\begin{itemize}

\item The re-annihilation of $D^+ D^-$ and $D^0 \bar D^0$ pairs into light
quarks will favor leading $ d \bar d$ and $u \bar u$ pairs, with amplitudes
in the ratio $d \bar d : u \bar u \simeq 2:3$ in line with the cross section
ratio $\sigma(D^+ D^-):\sigma(D^0 \bar D^0)$ (see the Appendix).  The
fragmentation of these quarks will populate hadronic final states in somewhat
different proportions than the usual continuum process in which quark pairs are
produced by the virtual photon with amplitudes proportional to their charges.

\item The re-annihilation largely favors isoscalar ($I=0$) odd-G-parity
final states, so one should see more effects of interference between
re-annihilation and continuum in odd G ($3 \pi,5 \pi, \eta 3 \pi, \eta' 3 \pi,
\ldots$) states than in even-G ones ($2 \pi, 4 \pi, \eta 2 \pi, \ldots$).
This interference is particularly pronounced because the larger odd-G
reannihilation amplitude is interfering with a smaller odd-G continuum
amplitude.

\item The effects of re-annihilation on the continuum contributions are quite
subtle if $\delta = 0$, especially in the dominant $I=1$ (even-G-parity)
channel.  They are proportionately greater in the $I=0$ (odd-G-parity)
non-strange channel (consisting, for example, of odd numbers of pions).

\item The re-annihilation may be similar to that which accounts for enhanced
penguin contributions in $B$ decays, particularly in the $b \to s$ subprocess
through the chain $b \to c \bar c s \to q \bar q s$, where $q=(u,d,s)$
(see also \cite{Rosner:2001nm}).  If this is so, one should look for an
enhancement of $\eta'$ production as appears to occur in both inclusive and
exclusive $B$ decays.

\item As especially evident for the examples of non-zero $\delta$, measurement
of the cross section in semi-inclusive channels with definite G-parity
and especially odd G (such as final states with an odd number of pions) may
show interesting interference patterns.

\end{itemize}

A Breit-Wigner amplitude is normally taken to be purely imaginary at its peak.
I incorporate this phase into the definition of $\delta$.  The choice $\delta =
3 \pi/2$ would correspond to no additional phase associated with
the re-annihilation process, for example in $e^+ e^- \to \mu^+ \mu^-$ in the
vicinity of the resonance, where interference between continuum and resonance
is destructive below the resonance and constructive above it.  (For an example
of this behavior at the $\psi'$, see Ref.\ \cite{Wang:2004qg}.)
It was speculated in Refs.\ \cite{Rosner:2001nm}
and \cite{Suzuki:fs} (see also Refs.\ \cite{Rosner:1999zm,Dun,Ciu,KLS})
that such an additional phase could be present and,
if related to a similar phase in $B$ decays, might account for a strong
phase in penguin $b \to s$ amplitudes.  A recent fit to $B \to P P$ decays,
where $P$ denotes a charmless pseudoscalar meson \cite{Chiang:2004nm}, finds
such a phase to be in the range of roughly $-20^\circ$ to $-50^\circ$.  This
would correspond to taking $\delta$ in the range of $40^\circ$ to $70^\circ$.
The presence of such a phase is supported by the recent strengthening of
the evidence for a significant CP asymmetry in the decay $B^0 \to K^+ \pi^-$
\cite{Aubert:2004qm}.

I now return briefly to a discussion of {\it specific} exclusive
charmless decay modes of the $\ppp$.  It was suggested in Ref.\
\cite{Rosner:2001nm} that some $\pp$ decay modes
might be suppressed via S--D mixing.  In that case, they should show up
in $\ppp$ decays.  Foremost among these was the $\pp \to \rho \pi$ decay.
A prediction was made for $\phi = (12 \pm 2)^\circ$ that $\Gamma(\ppp \to
\rho \pi) = (9.8 \pm 3.0)$ keV, corresponding to ${\cal B}(\ppp \to \rho
\pi) = (4.1 \pm 1.4) \times 10^{-4}$.
It was then pointed out \cite{Wang:2003zx} that because of possible
interference with continuum, decays such as $\ppp \to \rho \pi$ might
manifest themselves in various ways depending on relative strong phases, even
as a {\it dip} in $\sigma(e^+ e^- \to \rho \pi)$ at $M(\ppp)$.

An estimate of suppression of a $\pp$ decay rate may be constructed using
the quantity
\begin{equation}
Q(f) \equiv \frac
{{\cal B}(J/\psi \to e^+ e^-)}{{\cal B} (\psi' \to e^+ e^-)}
\frac{{\cal B}(\psi' \to f)}{{\cal B}(J/\psi \to f)}
\end{equation}
for any final state $f$. If $Q(f) < 1$, the decay $\pp \to f$ is suppressed
relative to $J/\psi \to f$, where the ratio of leptonic widths is an attempt
to correct for differing probabilities for $c \bar c$ annihilation.
Foremost among the $\pp$ modes which are candidates for some suppression is
$\pp \to \rho \pi$; this and several other modes have been tabulated,
for example, in Ref.\ \cite{FH}, based on a compilation of BES results.

The suppression mechanism is ascribed in Ref.\ \cite{Rosner:2001nm} to a
cancellation of the S- and D-wave contributions to $\pp \to f$:
\begin{equation}
\langle f|\psi' \rangle = \langle f|2^3S_1
\rangle \cos \phi - \langle f | 1^3D_1 \rangle \sin \phi = 0~,
\end{equation}
with a corresponding enhancement of the $\ppp \to f$ decay:
\begin{equation}
\langle f|\psi'' \rangle = \langle f|2^3S_1
\rangle \sin \phi + \langle f | 1^3D_1 \rangle \cos \phi = \langle f|2^3S_1
\rangle /\sin \phi~~.
\end{equation}
One can then use the predicted $\langle f|2^3S_1 \rangle$ matrix element and
the measured $\psi' \to f$ rate (whether an upper bound or observed) to
predict $\langle f|\psi'' \rangle$ and hence the $\ppp \to f$ rate.

All of the suppressed $\ppp$ modes discussed in Refs.\ \cite{Rosner:2001nm} and
\cite{FH} are prime candidates for detection in $\ppp$ decays.  However, the
interference proposed in Ref.\ \cite{Wang:2003zx} can actually lead to a {\it
suppression} of some modes relative to the rate expected from continuum.  A
firm conclusion will have to await more data both at the resonance and as a
function of c.m.\ energy in the neighboring continuum.  It was anticipated
in Ref.\ \cite{Rosner:2001nm} that if one were to account for any ``missing''
$\pp$ decay modes by mixing with the $\ppp$, such an effect need not
contribute more than a percent or two to the total $\ppp$ width.
\bigskip

\centerline{\bf VIII.  CONCLUSIONS}
\bigskip

Some non--$D \bar D$ decay modes of the $\psi''$ exist and are interesting in
their own right, such as $\ell^+ \ell^-$ pairs, $\gamma \chi_{cJ}$ and
$J/\psi \pi \pi$.  They tell us about mixing between S-waves, D-waves,
and open $D \bar D$ channels.

Most non--$D \bar D$
final states at the $\psi''$ are from continuum production.  Their yields
will not vary much with beam energy unless their continuum production
amplitudes are interfering with a genuine Breit-Wigner contribution from the
$\psi''$.  This interference is most likely to show up in odd-G-parity final
states, for which appreciable distortions of the Breit-Wigner line shape can
occur.

The suggestion that the ``missing'' $\psi'$ decays, like $\rho \pi$, should
show up instead at the $\psi''$, is being realized, if at all, in a more subtle
manner, and does not illuminate the question of whether a substantial fraction
(at least several percent) of the $\psi''$ cross section is non--$D \bar D$.
I predict a substantial enhancement of $\eta'$ production in charmless $\ppp$
final states if the re-annihilation of $D \bar D$ into light quarks is related
to the generation of a $b \to s$ penguin amplitude in $B$ decays.

The measurement of the continuum cross section at 3670 MeV is expected to
yield $R = 2 (1 + \alpha_S/\pi + \ldots)$.  Its value, when extrapolated to
3770 MeV, is relevant to whether there is a
cross section deficit at the $\psi''$.

Resolution of these questions is likely to require a measurement of the
$\psi''$ resonance shape, with an eye to possibly
different behavior in different channels.
\bigskip

\centerline{\bf ACKNOWLEDGMENTS}
\bigskip

I thank Karl Berkelman, David Cassel, Daniel Cronin-Hennessy, Richard Galik,
Brian Heltsley, Hanna Mahlke-Kr\"uger, Hajime Muramatsu, Ian Shipsey, Tomasz
Skwarnicki, and Misha Voloshin for discussions, and Maury Tigner for extending
the hospitality of the Laboratory for Elementary-Particle Physics at Cornell
during this research.  This work was supported in part by the United States
Department of Energy under Grant No.\ DE FG02 90ER40560, by the National
Science Foundation under Grant No.\ 0202078, and by the John Simon Guggenheim
Memorial Foundation.
\bigskip

\centerline{\bf APPENDIX:  MODEL FOR $D \bar D$ RE-ANNIHILATION}
\bigskip

The BES Collaboration's continuum value $R = 2.26 \pm 0.14$ \cite{Rong:2004ei}
averaged over 2 GeV $ \le E_{\rm c.m.} \le$ 3 GeV is consistent with the
expected value of 2 times a QCD correction (and also is consistent with the
background level obtained in the fit of Fig.\ 2 to the $\ppp$ cross section).
I take $R=2.26$ for illustration.  Of this, one expects $R(s \bar s) =
(1/6)(2.26) = 0.377$.  The non-strange contributions may be decomposed into a
9:1 ratio of $I=1$ and $I=0$ contributions denoted by $R_1$ and $R_0$, since
$(Q_u - Q_d)^2 = 9(Q_u + Q_d)^2$.  Thus $R_1 = (5/6)(2.26)(9/10) = 1.695$ and
$R_0 = (5/6)(2.26)(1/10) = 0.188$.
The $I=1$ continuum corresponds to an isovector photon and even-G-parity
states, while the $s \bar s$ and $I=0$ nonstrange continuua correspond to
an isoscalar photon and odd-G-parity states.  The $s \bar s$ continuum is
unlikely to lead to final states consisting exclusively of pions; one expects
at least one $K \bar K$ pair in its hadronic products.

A model of re-annihilation is to assume that the amplitude for $\ppp \to
D \bar D \to$ (non-charmed final states) proceeds via a $D \bar D$ loop diagram
characterized by an amplitude proportional to $(p^*)^3$, where $p^*$ is the
magnitude of the c.m.\ 3-momentum of either $D$.  For $\ppp \to D^+ D^-$,
$p^*_{+-} = 250.0$ MeV/$c$, while for $\ppp \to D^0 \od$, $p^*_{00} = 283.6$
MeV/$c$.  The re-annihilation amplitude $A^R_d$ into $d \bar d$ pairs and
the amplitude $A^R_u$ into $u \bar u$ pairs are then expected to be in the
ratio $A^R_d/A^R_u = (p_{+-}^*/p^*_{00})^3 = 0.685$, and the corresponding
ratio for isovector and nonstrange isoscalar contributions $A^R_1$ and $A^R_0$
is
\begin{equation}
\frac{A^R_1}{A^R_0} = \frac{A^R_u-A^R_d}{A^R_u+A^R_d}
= \frac{1 - 0.685}{1 + 0.685} = 0.187~~.
\end{equation}

One may assume for simplicity that the re-annihilation amplitudes into $I=0$
and $I=1$ final states have the same strong phase $\delta$ relative to the
continuum, modulated by a Breit-Wigner amplitude $f_B$ defined to be
unity at the resonance peak.  In the vicinity of the $\ppp$ mass $M_0$ one may
then write the amplitudes $A_1$ and $A_0$ for the isovector and nonstrange
isoscalar contributions to $R$ as functions of c.m.\ energy $E$:
\begin{equation}
A_1 = 0.187 b_0 e^{i \delta} f_B(E) + \sqrt{R_1}~~,~~~
A_0 = b_0 e^{i \delta} f_B(E) + \sqrt{R_0}~~,
\end{equation}
where the amplitudes have been defined such that their squares yield
their contributions to $R$, and
\begin{equation}
f_B(E) = [d_B(E)]^{-1}~~,~~~d_B(E) \equiv 1 + \frac{2 i (M_0 - E)}{\Gamma}~~.
\end{equation}
The values $M_0 = 3772$ MeV/$c^2$ and $\Gamma = 23.2$ MeV are taken from the
fit of Sec.\ VI.  This same fit implies a peak value $R(M_0) = 3.53$ which
will be taken as a constraint when choosing the arbitrary constant $b_0$.

The continuum away from the peak accounts for $R=2.26$, so one must provide
a total resonant contribution of $\Delta R_{\rm pk} = 3.53 - 2.26 = 1.27$.
For illustration, consider $D \bar D$ pairs to provide 82\% of this value,
or $\Delta R_{\rm pk}^{D \bar D} = 1.04$.  This contribution will be modulated
by $|f_B(E)|^2$.  There will be a constant $s \bar s$ continuum contribution
of $\Delta R^{s \bar s} = 0.38$, and contributions from the isovector and
non-strange isoscalar amplitudes $A_I$ above, leading to a total of
\begin{equation}
R(E) = |A_1|^2 + |A_0|^2 + \Delta R_{\rm pk}^{D \bar D}|f_B(E)|^2 +
\Delta R^{s \bar s}~~.
\end{equation}

For $\delta = 0$, a relatively modest value of $b_0 = 0.15$ provides
the additional contribution needed to account for the missing 18\% of the
$\ppp$ peak cross section.  The corresponding values for $\delta = \pi/2,
\pi,3\pi/2$ are 0.47, 1.46, and 0.47, respectively.  It is interesting
that the choice $\delta = \pi$, while implying large individual effects
in the $I=1$ and nonstrange $I=0$ channels, leads to an identical total
cross section shape when we demand $R(M_0) = 3.53$.  This result may
be demonstrated analytically with the help of the identity Re $f_B =
|f_B|^2$.

\bigskip

\def \ajp#1#2#3{Am.\ J. Phys.\ {\bf#1}, #2 (#3)}
\def \apny#1#2#3{Ann.\ Phys.\ (N.Y.) {\bf#1}, #2 (#3)}
\def \app#1#2#3{Acta Phys.\ Polonica {\bf#1}, #2 (#3)}
\def \arnps#1#2#3{Ann.\ Rev.\ Nucl.\ Part.\ Sci.\ {\bf#1}, #2 (#3)}
\def \b97{{\it Beauty '97}, Proceedings of the Fifth International
Workshop on $B$-Physics at Hadron Machines, Los Angeles, October 13--17,
1997, edited by P. Schlein}
\def \art{and references therein}
\def \cmts#1#2#3{Comments on Nucl.\ Part.\ Phys.\ {\bf#1}, #2 (#3)}
\def \cn{Collaboration}
\def \cp89{{\it CP Violation,} edited by C. Jarlskog (World Scientific,
Singapore, 1989)}
\def \ctp#1#2#3{Commun.\ Theor.\ Phys.\ {\bf#1}, #2 (#3)}
\def \efi{Enrico Fermi Institute Report No.\ }
\def \epjc#1#2#3{Eur.\ Phys.\ J. C {\bf#1}, #2 (#3)}
\def \f79{{\it Proceedings of the 1979 International Symposium on Lepton and
Photon Interactions at High Energies,} Fermilab, August 23-29, 1979, ed. by
T. B. W. Kirk and H. D. I. Abarbanel (Fermi National Accelerator Laboratory,
Batavia, IL, 1979}
\def \hb87{{\it Proceeding of the 1987 International Symposium on Lepton and
Photon Interactions at High Energies,} Hamburg, 1987, ed. by W. Bartel
and R. R\"uckl (Nucl.\ Phys.\ B, Proc.\ Suppl., vol.\ 3) (North-Holland,
Amsterdam, 1988)}
\def \ib{{\it ibid.}~}
\def \ibj#1#2#3{~{\bf#1}, #2 (#3)}
\def \ichep72{{\it Proceedings of the XVI International Conference on High
Energy Physics}, Chicago and Batavia, Illinois, Sept. 6 -- 13, 1972,
edited by J. D. Jackson, A. Roberts, and R. Donaldson (Fermilab, Batavia,
IL, 1972)}
\def \ijmpa#1#2#3{Int.\ J.\ Mod.\ Phys.\ A {\bf#1}, #2 (#3)}
\def \ite{{\it et al.}}
\def \jhep#1#2#3{JHEP {\bf#1}, #2 (#3)}
\def \jpb#1#2#3{J.\ Phys.\ B {\bf#1}, #2 (#3)}
\def \lg{{\it Proceedings of the XIXth International Symposium on
Lepton and Photon Interactions,} Stanford, California, August 9--14 1999,
edited by J. Jaros and M. Peskin (World Scientific, Singapore, 2000)}
\def \lkl87{{\it Selected Topics in Electroweak Interactions} (Proceedings of
the Second Lake Louise Institute on New Frontiers in Particle Physics, 15 --
21 February, 1987), edited by J. M. Cameron \ite~(World Scientific, Singapore,
1987)}
\def \kdvs#1#2#3{{Kong.\ Danske Vid.\ Selsk., Matt-fys.\ Medd.} {\bf #1},
No.\ #2 (#3)}
\def \ky85{{\it Proceedings of the International Symposium on Lepton and
Photon Interactions at High Energy,} Kyoto, Aug.~19-24, 1985, edited by M.
Konuma and K. Takahashi (Kyoto Univ., Kyoto, 1985)}
\def \mpla#1#2#3{Mod.\ Phys.\ Lett.\ A {\bf#1}, #2 (#3)}
\def \nat#1#2#3{Nature {\bf#1}, #2 (#3)}
\def \nc#1#2#3{Nuovo Cim.\ {\bf#1}, #2 (#3)}
\def \nima#1#2#3{Nucl.\ Instr.\ Meth. A {\bf#1}, #2 (#3)}
\def \np#1#2#3{Nucl.\ Phys.\ {\bf#1}, #2 (#3)}
\def \npbps#1#2#3{Nucl.\ Phys.\ B Proc.\ Suppl.\ {\bf#1}, #2 (#3)}
\def \os{XXX International Conference on High Energy Physics, Osaka, Japan,
July 27 -- August 2, 2000}
\def \PDG{Particle Data Group, K. Hagiwara \ite, \prd{66}{1}{2002}}
\def \pisma#1#2#3#4{Pis'ma Zh.\ Eksp.\ Teor.\ Fiz.\ {\bf#1}, #2 (#3) [JETP
Lett.\ {\bf#1}, #4 (#3)]}
\def \pl#1#2#3{Phys.\ Lett.\ {\bf#1}, #2 (#3)}
\def \pla#1#2#3{Phys.\ Lett.\ A {\bf#1}, #2 (#3)}
\def \plb#1#2#3{Phys.\ Lett.\ B {\bf#1}, #2 (#3)}
\def \pr#1#2#3{Phys.\ Rev.\ {\bf#1}, #2 (#3)}
\def \prc#1#2#3{Phys.\ Rev.\ C {\bf#1}, #2 (#3)}
\def \prd#1#2#3{Phys.\ Rev.\ D {\bf#1}, #2 (#3)}
\def \prl#1#2#3{Phys.\ Rev.\ Lett.\ {\bf#1}, #2 (#3)}
\def \prp#1#2#3{Phys.\ Rep.\ {\bf#1}, #2 (#3)}
\def \ptp#1#2#3{Prog.\ Theor.\ Phys.\ {\bf#1}, #2 (#3)}
\def \rmp#1#2#3{Rev.\ Mod.\ Phys.\ {\bf#1}, #2 (#3)}
\def \rp#1{~~~~~\ldots\ldots{\rm rp~}{#1}~~~~~}
\def \rpp#1#2#3{Rep.\ Prog.\ Phys.\ {\bf#1}, #2 (#3)}
\def \sing{{\it Proceedings of the 25th International Conference on High Energy
Physics, Singapore, Aug. 2--8, 1990}, edited by. K. K. Phua and Y. Yamaguchi
(Southeast Asia Physics Association, 1991)}
\def \slc87{{\it Proceedings of the Salt Lake City Meeting} (Division of
Particles and Fields, American Physical Society, Salt Lake City, Utah, 1987),
ed. by C. DeTar and J. S. Ball (World Scientific, Singapore, 1987)}
\def \slac89{{\it Proceedings of the XIVth International Symposium on
Lepton and Photon Interactions,} Stanford, California, 1989, edited by M.
Riordan (World Scientific, Singapore, 1990)}
\def \smass82{{\it Proceedings of the 1982 DPF Summer Study on Elementary
Particle Physics and Future Facilities}, Snowmass, Colorado, edited by R.
Donaldson, R. Gustafson, and F. Paige (World Scientific, Singapore, 1982)}
\def \smass90{{\it Research Directions for the Decade} (Proceedings of the
1990 Summer Study on High Energy Physics, June 25--July 13, Snowmass, Colorado),
edited by E. L. Berger (World Scientific, Singapore, 1992)}
\def \tasi{{\it Testing the Standard Model} (Proceedings of the 1990
Theoretical Advanced Study Institute in Elementary Particle Physics, Boulder,
Colorado, 3--27 June, 1990), edited by M. Cveti\v{c} and P. Langacker
(World Scientific, Singapore, 1991)}
\def \yaf#1#2#3#4{Yad.\ Fiz.\ {\bf#1}, #2 (#3) [Sov.\ J.\ Nucl.\ Phys.\
{\bf #1}, #4 (#3)]}
\def \zhetf#1#2#3#4#5#6{Zh.\ Eksp.\ Teor.\ Fiz.\ {\bf #1}, #2 (#3) [Sov.\
Phys.\ - JETP {\bf #4}, #5 (#6)]}
\def \zpc#1#2#3{Zeit.\ Phys.\ C {\bf#1}, #2 (#3)}
\def \zpd#1#2#3{Zeit.\ Phys.\ D {\bf#1}, #2 (#3)}

\end{document}